\documentstyle[12pt]{article}

\begin{document}
\setcounter{page}{1} \pagestyle{plain} \vspace{1cm}
\begin{center}
\Large{\bf Quantum-Corrected Black Hole Thermodynamics in Extra Dimensions}\\
\small
\vspace{1cm} {\bf Kourosh Nozari}\quad and \quad {\bf S. Hamid Mehdipour}\\
\vspace{0.5cm} {\it Department of Physics,
Faculty of Basic Sciences,\\
University of Mazandaran,\\
P. O. Box 47416-1467,
Babolsar, IRAN}\\
{\it e-mail: knozari@umz.ac.ir}

\end{center}
\vspace{1.5cm}

\begin{abstract}
Bekenstein-Hawking formalism of black hole thermodynamics should be
modified to incorporate quantum gravitational effects. Generalized
Uncertainty Principle(GUP) provides a suitable framework to perform
such modifications. In this paper, we consider a general form of GUP
to find black hole thermodynamics in a model universe with large
extra dimensions. We will show that black holes radiate mainly in
the four-dimensional brane. Existence of black holes remnants
as a possible candidate for dark matter is discussed.
\end{abstract}
\newpage

\section{Introduction}
The idea of Large Extra Dimensions (LEDs) which recently has been
proposed [1-5], might allow to study interactions at trans-Planckian
energies in the next generation collider experiments. The ADD-model
proposed by Arkani-Hamed, Dimopoulos and Dvali[1-3] adds $d$ extra
spacelike dimensions without curvature, in general each of them
compactified to the same radius $L$. In this scenario, all
standard-model particles are confined to the observable
4-dimensional brane universe, whereas gravitons can access the whole
$d$-dimensional bulk spacetime, being localized at the brane at low
energies. In this scenario, the hierarchy problem is solved or at
least reformulated in a geometric language. On the other hand, the
setting of RS-model proposed by Randall and Sundrum[4,5] is a
5-dimensional spacetime with an non-factorizable geometry. The
solution for the metric is found by analyzing the solution of
Einstein's field equations with an energy density on our brane,
where the standard model particles live. In the type I model the
extra dimension is compactified while in the type II model it is
infinite.\\ The possibility of the existence of large extra
dimensions has opened up new and exciting avenues of research in
quantum gravity. In particular, a host of interesting work is being
done on different aspects of low-energy scale quantum gravity
phenomenology. One of the most significant sub-fields is the study
of black hole production at particle colliders, such as the Large
Hadronic Collider (LHC)[6] and the muon collider [7], as well as in
ultrahigh energy cosmic ray (UHECR) airshowers [8,9]. Newly formed
black holes first lose hair associated with multipole and angular
momenta, then approach classically stable Schwarzschild solutions,
and finally evaporate via Hawking radiation [10] up to  possible
Planck size remnants. Decay time and entropy completely determine
the observables of the process. Black hole formation and decay can
be described semiclassically, provided that the entropy is
sufficiently large. The timescale for the complete decay of a black
hole up to its supposed final Planck-sized remnant is expected to be
of order of the $TeV^{-1}$. Black Hole thermodynamical quantities
depend on the Hawking temperature $T_H$ via the usual
thermodynamical relations (for example Stefan-Boltzmann law). The
Hawking temperature undergoes corrections from many sources, and
these corrections are particularly relevant for black holes with
mass of the order of the Planck mass. Therefore, the study of
$TeV$-scale black holes in UHECR and particle colliders requires a
careful investigation of how temperature corrections affect black
hole thermodynamics. In this article, we concentrate on the
corrections due to the generalized uncertainty principle (GUP) in
the framework of LEDs. These corrections are not tied down to any
specific model of quantum gravity; since GUP can be derived using
arguments from string theory [11] as well as other approaches to
quantum gravity [12,13]. Black holes thermodynamics in four
spacetime dimensions and in the framework of GUP, has been studied
in several context[14-18]. Embedding a black hole in a space-time of
higher dimensionality would seem, from the string theory point of
view, to be a natural thing to do. Black holes in $d$ extra
dimensions have been studied in both compact [19] and infinitely
extended [20] extra dimensions (see also [21] and references
therein). Here we proceed one more step in this direction. Using a
general form of GUP, we provide a perturbational framework to
calculate temperature and entropy of a black hole in a model
universe with large extra dimensions. Our approach will show that
black holes decay mainly on the brane. We investigate also the
possibility of having black holes remnants in extra dimensional
scenarios. These remnants are good candidates for dark
matter.\\
The paper is organized as follows: Section 2 gives our primary
inputs for rest of the calculations. Section 3 is devoted to
calculation of GUP-induced corrections of black hole thermodynamics.
Section 4 considers the black holes remnants as a possible source of
dark matter. The paper follows by conclusions in section 5.\\

\section{GUP and LEDs}
The canonical commutation relations between the momentum operator
$p^{\nu}$ and position operator $x^{\mu}$, which in Minkowski
space-time are $[x^{\mu}, p^{\nu}] = i\hbar \eta^{\mu\nu}$, in a
curved space-time with metric $g_{\mu\nu}$  can be generalized as
\begin{equation}
\label{math:3.1} [x^{\mu}, p^{\nu}] = i\hbar g^{\mu\nu}(x).
\end{equation}
This equation contains gravitational effects of a particle in first
quantization scheme. Its validity is confined to curved spacetime
asymptotically flat so that the tensor metric can be decomposed as
$g_{\mu\nu}= \eta_{\mu\nu} + h_{\mu\nu}$, where $h_{\mu\nu}$ is the
local perturbation to the flat background[22]. We note that the
usual commutation relations between position and momentum operators
in Minkowsky spacetime are obtained by using the veirbein formalism,
i.e. by projecting the commutator and the metric tensor on the
tangent space. In which follows we consider another alternative:
existence of a minimal observable length. As it is well known, a
theory containing a fundamental length on the order of $l_P$ (which
can be related to the extension of particles) is string theory. It
provides a consistent theory of quantum gravity and allows to avoid
the above mentioned difficulties. In fact, unlike point particle
theories, the existence of a fundamental length plays the role of
natural cut-off. In such a way the ultraviolet divergencies are
avoided without appealing to the renormalization and regularization
schemes[23]. Besides, by studying string collisions at Planckian
energies and through a renormalization group type analysis the
emergence of a minimal observable distance yields to the generalized
uncertainty principle
\begin{equation}
\label{math:3.2}\delta x\geq\frac{\hbar}{2\delta p}+const.G\delta p,
\end{equation}
At energy much below the Planck mass, $m_{p} = \sqrt{\frac{\hbar
c}{G}} \sim 10^{19} GeV/c^{2}$, the extra term in equation (2) is
irrelevant and the Heisenberg uncertainty relation is recovered,
while, as we approach the Planck energy, this term becomes relevant
and is related to the minimal observable length on the order of
Planck length,
$l_p=\sqrt{\frac{G\hbar}{c^3}}\sim1.6\times10^{-35}m$. In terms of
Planck length, equation (2) can be written as,
\begin{equation}
\label{math:3.2}\delta x\geq\frac{\hbar}{2}\bigg(\frac{1}{\delta
p}+\alpha^2 l_{p}^{2}\frac{\delta p}{\hbar}\bigg).
\end{equation}
where $\alpha$ is a dimensionless constant of order one which
depends on the details of the quantum gravity theory. Note that one
should consider an extra term in the right hand side of relation (3)
which depends to expectation values of $x$ and $p$. But since we
like to dealing with absolutely smallest position uncertainty, this
extra term is omitted.\\
It is important to note that various domains of modern physics lead
to such a result. In addition to string theory, theories such as
loop quantum gravity, black hole gedanken experiments and quantum
geometry also give such generalized uncertainty principle. The
matter which is very interesting is the fact that these generalized
uncertainty principles can be obtained in the framework of classical
Newtonian gravitational theory and classical general relativity[24].
In other words, although a full description of quantum gravity is
not yet available, there are some general features that seem to go
hand in hand with all promising candidates for such a theory where
one of them is the existence of a minimal length scale. This minimal
length scale gives an extreme quantum regime called Planck regime.
In this scale the running couplings unify and quantum gravity era is
likely to occur. At this scale the quantum effects of gravitation
get as important as those of the electroweak and strong
interactions. In this extreme conditions, the usual Heisenberg
algebra should be modified regarding extra uncertainty induced by
quantum gravitational effects. This modified commutator algebra may
be given as follows
\begin{equation}
\label{math:2.3} [x,p]=i\hbar(1+\beta' p^2),
\end{equation}
It is possible to have more terms in the right hand side of (2). One
can consider more generalized uncertainty relation such as[25],
\begin{equation}
\label{math:2.4} \delta x\delta
p\geq\frac{\hbar}{2}\Big(1+\alpha'(\delta x)^2+\beta'(\delta
p)^2+\gamma\Big),
\end{equation}
which leads to a nonzero minimal uncertainty in both position and
momentum. This relation can lead us to following commutator relation
\begin{equation}
\label{math:2.5}[x,p]=i\hbar\Big(1+\alpha' x^2+\beta' p^2\Big),
\end{equation}
where $\gamma = \alpha'\langle x\rangle^{2} + \beta'\langle
p\rangle^{2}$. This statement shows that GUP itself has a
perturbational expansion. We are going to consider the effects of
GUP on black hole thermodynamics in model universes with extra
dimensions. Therefore we need to representation of GUP in LEDs
scenarios. As has been indicated, there are two main scenarios of
extra dimensions:
\begin{itemize}
\item
the Arkani-Hamed--Dimopoulos--Dvali (ADD) model[1-3], where
the extra dimensions are compact and of size $L$;\\
and
\item
the Randall--Sundrum (RS) model[4,5], where the extra dimensions
have an infinite extension but are warped by a non-vanishing
cosmological constant.
\end{itemize}
A feature shared by (the original formulations of) both scenarios is
that only gravity propagates along the extra dimensions, while
Standard Model fields are confined on a four-dimensional
sub-manifold usually referred to as the  brane-world. In which
follows we consider the ADD model as our LEDs scenario. In LEDs
scenario, GUP can be written as follows
\begin{equation}
\label{math:2.5} \delta x_i\delta p_i\geq
\frac{\hbar}{2}\bigg(1+\frac{{\alpha}^2 L^2_{Pl}}{\hbar^2}(\delta
p_i)^2+\frac{{\beta}^2}{L^2_{Pl}}(\delta x_i)^2+\gamma\bigg).
\end{equation}
Here $\alpha$, $\beta$ and $\gamma$ are dimensionless, positive and
independent of $\delta x$ and $\delta p$ but may in general depend
on the expectation values of $x$ and $p$. Planck length now is
defined as $ L_{Pl}= \Big(\frac{\hbar
G_d}{c^{3}}\Big)^{\frac{1}{d-2}}$. Here $G_d$ is gravitational
constant in $d$ dimensional spacetime which in ADD model is given by
$G_{d} = G_{4}L^{n}$ where $n$ is number of extra dimensions, $n=d-4$.\\
In which follows, we use this more general form of GUP as our
primary input and construct a perturbational calculations to find
thermodynamical properties of black hole and its quantum
gravitational corrections. It should be noted that since GUP is a
model independent concept[26], the results which we obtain are
consistent with any fundamental theory of
quantum gravity.\\

\section{Black Holes Thermodynamics}
\subsection{Black Holes Temperature}
The Hawking temperature for a spherically symmetric black hole may
be obtained in a heuristic way with the use of the standard
uncertainty principle and general properties of black holes [27]. We
picture the quantum vacuum as a fluctuating sea of virtual
particles; the virtual particles cannot normally be directly
observed without violating energy conservation. But near the surface
of a black hole the effective potential energy can negate the rest
energy of a particle and give it zero total energy, and the surface
itself is a one-way membrane which can swallow particles so that
they are henceforth not observable from outside. The net effect is
that for a pair of photons one photon may be absorbed by the black
hole with effective negative energy $-E$, and the other may be
emitted to asymptotic distances with positive energy $+E$. The
characteristic energy $E$ of the emitted photons may be estimated
from the standard uncertainty principle. In the vicinity of the
black hole surface there is an intrinsic uncertainty in the position
of any particle of about the Schwarzschild radius, $r_s$ , due to
the behavior of its field lines [28], as well as on dimensional
grounds. This leads to momentum uncertainty
\begin{equation}
\label{math:3.1} \delta p\approx\frac{\hbar}{\delta
x}=\frac{\hbar}{r_s}=\frac{\hbar c^2}{2GM},\qquad \delta x\approx
r_s=\frac{2GM}{c^2}
\end{equation}
and to an energy uncertainty of $\delta pc = \frac{\hbar c^3}{
2GM}$. We identify this as the characteristic energy of the emitted
photon, and thus as a characteristic temperature; it agrees with the
Hawking temperature up to a factor of $4\pi $, which we will
henceforth include as a "calibration factor" and write, with $k_B =
1$,
\begin{equation}
\label{math:3.1} T_{H}\approx \frac{\hbar c^3}{8\pi GM},
\end{equation}
 The related entropy is obtained by
integration of $dS_B=\frac{c^2dM}{T_H}$ which is the standard
Bekenstein entropy,
\begin{equation}
\label{math:3.2} S_{B}=\frac{4\pi GM^2}{\hbar c}= \frac{A}{4l_p^2}
\end{equation}
where $A=4\pi r_s^2$(the area of event horizon ). A $d$-dimensional
spherically symmetric BH of mass $M$ (to which the collider BHs will
settle into before radiating) is described by the metric,
\begin{equation}
\label{math:3.4}ds^2=-\bigg(1-\frac{16\pi
G_dM}{(d-2)\Omega_{d-2}c^2r^{d-3}}\bigg)c^2dt^2+\bigg(1-\frac{16\pi
G_dM}{(d-2)\Omega_{d-2}c^2r^{d-3}}\bigg)^{-1}dr^2+r^2d\Omega^2_{d-2}
\end{equation}
where $\Omega_{d-2}$ is the metric of the unit $S^{d-2}$ as
$\Omega_{d-2}=\frac{2\pi^{\frac{d-1}{2}}}{\Gamma(\frac{d-1}{2})}$.
Since the Hawking radiation is a quantum process, the emitted quanta
should satisfy the generalized uncertainty principle(which has
quantum gravitational nature) in its general form. Therefore, we
consider equation (7), where $x_i$ and $p_i$ with $i = 1 . . . d -
1$, are the spatial coordinates and momenta respectively. By
modeling a BH as a $(d-1)$-dimensional cube of size equal to its
Schwarzschild radius $r_s$, the uncertainty in the position of a
Hawking particle at the emission is,
\begin{equation}
\label{math:3.4}\delta x_i\approx
r_s=\omega_dL_{Pl}m^{\frac{1}{d-3}},
\end{equation}
where $$ \omega_d=\bigg(\frac{16\pi}{(d-2)\Omega_{d-2}}\bigg)
^{\frac{1}{d-3}},$$ $m=\frac{M}{M_{Pl}}$ and
$M_{Pl}=\bigg(\frac{\hbar^{d-3}}{c^{d-5}G_d}\bigg)^{\frac{1}{d-2}}$.
Here $\omega_d$ is dimensionless area factor. A simple calculation
based on equation (7) gives,
\begin{equation}
\label{math:3.1} \delta x_i\simeq\frac{L_{Pl}^2 \delta
p_i}{{\beta}^2 \hbar}\Bigg[1\pm\sqrt{1-{\beta}^2\bigg(
{\alpha}^2+\frac{\hbar^2(\gamma+1)}{L_{Pl}^2(\delta
p_i)^2}\bigg)}\,\,\Bigg].
\end{equation}
Here, to achieve standard values (for example $\delta x_i\delta
p_i\geq \hbar$) in the limit of $\alpha=\beta=\gamma =0$, we should
consider the minus sign. One can minimize $\delta x$ to find
\begin{equation}
\label{math:3.2}(\delta x_i)_{min}\simeq\pm \alpha
L_{Pl}\sqrt{\frac{1+\gamma}{1-\alpha^2\beta^2}}.
\end{equation}
This is minimal observable length on the order of Planck length.
Here we should consider the plus sign in equation (14), whereas the
negative sign has no evident physical meaning. Equation (7) gives
also
\begin{equation}
\label{math:3.3} \delta p_i\simeq\frac{\hbar \delta
x_i}{{\alpha}^2L_{Pl}^2}\Bigg[1\pm\sqrt{1-{\alpha}^2\bigg({\beta}^2+\frac{
L_{Pl}^2(\gamma+1)}{(\delta x_i)^2}\bigg)}\Bigg].
\end{equation}
Here to achieve correct limiting results we should consider the
minus sign in round bracket. From a heuristic argument based on
Heisenberg uncertainty relation, one deduces the following equation
for Hawking temperature of black holes[14],
\begin{equation}
\label{math:3.4}T_H\approx \frac{(d-3)c\delta p_i }{4\pi}
\end{equation}
where we have set the constant of proportionality equal to
$\frac{(d-3)}{4\pi}$ in extra dimensional scenarios. Based on this
viewpoint, but now using generalized uncertainty principle in its
general form, modified black hole temperature in GUP is,
\begin{equation}
\label{math:3.5}T^{GUP}_{H}\approx \frac{(d-3)\hbar c \delta
x_i}{{4\pi \alpha}^2 L^2_{Pl}}\Bigg[1-\sqrt{1-{\alpha}^2\bigg(
{\beta}^2+\frac{L^2_{Pl}(\gamma+1)}{(\delta
x_i)^2}\bigg)}\,\,\Bigg].
\end{equation}
Since $\delta x_{i}$ is given by (12), this relation can be
expressed in terms of black hole mass in any stage of its
evaporation. Figure 1 shows the relation between temperature and
mass of the black hole in different spacetime dimensions. Following
results can be obtained from this analysis : In scenarios with extra
dimensions, black hole temperature increases. This feature leads to
faster decay and less classical behaviors for black holes. It is
evident that in extra dimensional scenarios final stage of
evaporation( black hole remnant) has mass more than its four
dimensional counterpart. Therefore, in the framework of GUP, it
seems that quantum black holes are hotter, shorter-lived and tend to
evaporate less than classical black holes. Note that these results
are applicable to both ADD and RS brane world scenarios.\\
\subsection{Black Holes Entropy}
Now consider a quantum particle that starts out in the vicinity of
an event horizon and then ultimately absorbed by black hole. For a
black hole absorbing such a particle with energy $E$ and size $l$,
the minimal increase in the horizon area can be expressed as [29]
\begin{equation}
\label{math:3.6}(\Delta \textsf{A})_{min}\geq\frac{ 8\pi
L_{Pl}^{d-2}E l}{(d-3)\hbar c},
\end{equation}
then one can write
\begin{equation}
\label{math:3.7}(\Delta \textsf{A})_{min}\geq\frac{ 8\pi
L_{Pl}^{d-2}c\delta p_i l}{(d-3)\hbar c},
\end{equation}
where $E\sim c\delta p_i $ and  $l\sim\delta x_i$.
\begin{equation}
\label{math:3.8} (\Delta \textsf{A})_{min}\simeq\frac{8\pi
L_{Pl}^{d-4}(\delta
x_i)^2}{(d-3){\alpha}^2}\Bigg[1-\sqrt{1-{\alpha}^2\bigg({\beta}^2+\frac{
L_{Pl}^2(\gamma+1)}{(\delta x_i)^2}\bigg)}\Bigg],
\end{equation}
Now we should determine $\delta x_i$. Since our goal is to compute
microcanonical entropy of a large black hole, near-horizon geometry
considerations suggests the use of inverse surface gravity or simply
the Schwarzschild radius for $\delta x_i$. Therefore, $\delta
x_i\approx r_s$ and defining $\Omega_{d-2} r_s^{d-2}=\textsf{A}$ or
$r_s^2=\Omega_{d-2}^{-\frac{2}{d-2}}\textsf{A}^{\frac{2}{d-2}}$ and
$(\Delta S)_{min}=b$, then it is easy to show that,
\begin{equation}
\label{math:3.8} (\Delta \textsf{A})_{min}\simeq\frac{8\pi
L_{Pl}^{d-4}\Omega_{d-2}^{-\frac{2}{d-2}}\textsf{A}^{\frac{2}{d-2}}}{(d-3){\alpha}^2}
\Bigg[1-\sqrt{1-{\alpha}^2\bigg({\beta}^2+\frac{
L_{Pl}^2(\gamma+1)}{\Omega_{d-2}^{-\frac{2}{d-2}}\textsf{A}^{\frac{2}{d-2}}}\bigg)}\Bigg],
\end{equation}
and,
\begin{equation}
\label{math:3.9}\frac{dS}{d\textsf{A}}\simeq\frac{(\Delta
S)_{min}}{(\Delta
\textsf{A})_{min}}\simeq\frac{\Omega_{d-2}^{\frac{2}{d-2}}b{\alpha}^2(d-3)}{8\pi
L_{Pl}^{d-4}\textsf{A}^{\frac{2}{d-2}}\Bigg[1-\sqrt{1-{\alpha}^2\bigg({\beta}^2+\frac{
\Omega_{d-2}^{\frac{2}{d-2}}L_{Pl}^2(\gamma+1)}{\textsf{A}^{\frac{2}{d-2}}}\bigg)}\Bigg]}.
\end{equation}
Two point should be considered here. First note that $b$ can be
considered as one bit of information since entropy is an extensive
quantity. Secondly, in our approach we consider microcanonical
ensemble since we are dealing with Schwarzschild black hole of fixed
mass. Now we should perform integration. There are two possible
choices for lower limit of integration, $\textsf{A}=0$ and
$\textsf{A}=\textsf{A}_p$ . Existence of a minimal observable length
leads to existence of a minimum event horizon area, $\textsf{A}_p =
\Omega_{d-2} (\delta x_i)_{min}^{d-2}$. So it is physically
reasonable to set $\textsf{A}_p$ as lower limit of integration. This
is in accordance with existing picture[14]. Based on these
arguments, we can write
\begin{equation}
\label{math:3.7}S\simeq\varepsilon\int_{\textsf{A}_{p}}^\textsf{A}\frac{\textsf{A}^{-\frac{2}{d-2}}}{1-\sqrt{\eta+\kappa
\textsf{A}^{-\frac{2}{d-2}}}}d\textsf{A}
\end{equation}
where, $$\varepsilon=
\frac{\Omega_{d-2}^{\frac{2}{d-2}}b\alpha^2(d-3)}{8\pi L_{Pl}^{d-4}
},\quad\quad \kappa=-{\Omega_{d-2}^{\frac{2}{d-2}}\alpha}^{2}
L_{Pl}^{2}(\gamma+1), \quad \quad \eta= 1-{\alpha}^{2}{\beta}^{2},$$
\begin{equation}
\label{math:3.7}\textsf{A}_{p}=\Omega_{d-2}\big(\alpha
L_{Pl}\big)^{d-2}\Bigg(\frac{1+\gamma}
{1-\alpha^2\beta^2}\Bigg)^{\frac{(d-2)}{2}}
\end{equation}
This integral can be solved numerically. The result is shown in
figure 2. This figure shows that: In scenarios with extra
dimensions, black hole entropy decreases. The classical picture
breaks down since the degrees of freedom of the black hole, i.e. its
entropy, is small. In this situation one can use the semiclassical
entropy to measure the validity of the semiclassical approximation.
It is evident that in extra dimensional scenarios final stage of
evaporation( black hole remnant) has event horizon area more than
its four dimensional counterpart. Therefore, higher dimensional
black hole remnants have less classical features relative to their
four dimensional counterparts. To obtain the relation between
emission rate of black holes radiation and spacetime dimensions, we
proceed as follows. As Emparan {\it et al} have shown[21], in $d$
dimensions, the energy radiated by a black body of temperature $T$
and surface area $\textsf{A}$  is given by
\begin{equation}
\frac{dE_{d}}{dt}=\sigma_{d} \textsf{A} T^{d},
\end{equation}
where $\sigma_{d}$ is $d$-dimensional Stefan-Boltzman constant,
$$\sigma_{d} = \frac{\Omega_{d-3}}{(2\pi)^{d-1}(d-2)}\Gamma(d)
\zeta(d).$$ Now using equations (17) for modified Hawking
temperature in the framework of GUP, equation (25) becomes
\begin{equation}
\frac{dE_{d}}{dt}=
\frac{\Omega_{d-3}\Omega_{d-2}}{(2\pi)^{d-1}(d-2)}\Gamma(d)
\zeta(d)\Bigg(\frac{(d-3)\hbar c}{{4\pi \alpha}^2
L^2_{Pl}}\Bigg[1-\sqrt{1-{\alpha}^2\bigg(
{\beta}^2+\frac{L^2_{Pl}(\gamma+1)}{(r_{s})^2}\bigg)}\,\,\Bigg]\Bigg)^d
r_s^{2d-2},
\end{equation}
where we have set $\delta x_{i}\sim r_{s}$. This is a complicated
relation. To compare emission rates of black holes in different $d$,
note that $\sigma_{n}$ changes very little with dimension. This fact
confirms that even though higher dimensional spacetimes have
infinitely many more modes due to excitations in the extra
dimensions, the rate at which energy is radiated by black body with
radius $r_{s}$ and temperature $T\sim\frac{1}{r_{s}}$ is roughly
independent of the dimension. Based on this argument, let us assume
that $d=4$, $d=6$ and $d=10$. Since $\sigma_{4} =0.08$, $\sigma_{6}
=0.06$ and $\sigma_{10} =0.097$, some numerical calculations give
approximately
\begin{equation}
\frac{(\frac{dE_{4}}{dt})}{(\frac{dE_{6}}{dt})}\approx11\quad\quad\quad
and\quad\quad\quad
\frac{(\frac{dE_{4}}{dt})}{(\frac{dE_{10}}{dt})}\approx 12
\end{equation}
These results evidently show that black holes radiate mainly on the
$4$-dimensional brane. In fact, a higher-dimensional black hole
emits radiation both in the bulk and on the brane. Note that some
corrections to equation (27) should be considered due to area
appeared in (25). A detailed calculation shows that critical radius
of black hole as an absorber is given by[21]
\begin{equation}
r_{c}=\Bigg(\frac{d-1}{2}\Bigg)^{\frac{1}{d-3}}
\Bigg(\frac{d-1}{d-3}\Bigg)^{1/2}r_{s}.
\end{equation}
Therefore, equation (27) will change to
\begin{equation}
\frac{(\frac{dE_{4}}{dt})}{(\frac{dE_{6}}{dt})}\approx3.5\quad\quad\quad
and\quad\quad\quad
\frac{(\frac{dE_{4}}{dt})}{(\frac{dE_{10}}{dt})}\approx 1.5
\end{equation}
According to the assumptions of the theory with Large Extra
Dimensions, only gravitons, and possibly scalar fields, can
propagate in the bulk and thus, these are the only types of fields
allowed to be emitted in the bulk during the Hawking evaporation
phase. On the other hand, the emission on the brane can take the
form of scalar Higgs particles, fermions and gauge bosons. From the
perspective of the brane observer, the radiation emitted in the bulk
will be a missing energy signal, while radiation on the brane may
lead to experimental detection of Hawking radiation and thus of the
production of small black holes. In next section we discuss some of
these experimental
approaches.\\

\section{Black Holes Remnants and Extra Dimensions}
It is by now widely accepted that dark matter (DM) constitutes a
substantial fraction of the present critical energy density in the
Universe. However, the nature of DM remains an open problem. There
exist many DM candidates, most of them are non-baryonic weakly
interacting massive particles (WIMPs), or WIMP-like particles[30].
By far the DM candidates that have been more intensively studied are
the lightest supersymmetric (SUSY) particles such as neutralinos or
gravitinos, and the axions (as well as the axinos). There are
additional particle physics inspired dark matter candidates[30]. A
candidate which is not as closely related to particle physics is the
relics of primordial black holes(Micro Black Holes)[31,32]. Certain
inflation models naturally induce a large number of such a black
holes. As a specific example, hybrid inflation can in principle
yield the necessary abundance of primordial black hole remnants for
them to be the primary source of dark matter[33,34]. Here we have
shown that final stage of evaporation of a black hole is a remnant
which has mass increasing with spacetime dimensions. One of the
major problems with these remnants is the possibility of their
detection. As interactions with black hole remnants are purely
gravitational, the cross section is extremely small, and direct
observation of these remnants seems unlikely. One possible indirect
signature may be associated with the cosmic gravitational wave
background. Unlike photons, the gravitons radiated during
evaporation would be instantly frozen. Since, according to our
notion, the black hole evaporation would terminate when it reduces
to a remnants, the graviton spectrum should have a cutoff at Planck
mass. Such a cutoff would have by now been red-shifted to
$\sim10^{14} GeV$. Another possible gravitational wave-related
signature may be the gravitational wave released during the
gravitational collapse. The frequencies of such gravitational waves
would by now be in the range of $\sim 10^{7} - 10^{8} Hz$. It would
be interesting to investigate whether these signals are in principle
observable. Another possible signature may be some imprints on the
cosmic microwave background(CMB) fluctuations due to the
thermodynamics of black hole remnants-CMB interactions. Possible
production of such remnants in Large Hadron Collider (LHC) and also
in ultrahigh energy cosmic ray (UHECR) air showers are under
investigation. If we consider hybrid inflation as our primary
cosmological model, there will be some observational constraints on
hybrid inflation parameters. For example a simple calculation based
on hybrid inflation suggests that the time it took for black holes
to reduce to remnants is about $10^{-10} Sec$. Thus primordial black
holes have been produced before baryogenesis and subsequent epochs
in the standard cosmology[35]. The events that can potentially lead
to black hole production are essentially high-energy scattering in
particle colliders and UHECR. The next generation of particle
colliders are expected to reach energies above $10$ $TeV$. LHC and
Very Large Hadron Collider (VLHC)[36] are planned to reach a
center-of-mass energy of $14$ and $100$ $TeV$. Therefore, if the
fundamental Planck scale is of the order of few $TeV$, LHC and VLHC
would copiously produce black holes. These black holes have masses
on the order of $TeV$.\\
Black hole production by cosmic rays has also been recently
investigated by a number of authors[37]. Cosmogenic neutrinos[38]
with energies above the Greisen- Zatsepin-Kuzmin (GZK) cutoff[39]
are expected to create black holes in the terrestrial atmosphere.
The thermal decay of the black hole produces air showers which could
be observed. The cross sections of these events are two or more
orders of magnitude larger than the cross sections of standard model
processes. Therefore, black holes are created uniformly at all
atmospheric depths with the most promising signal given by
quasi-horizontal showers which maximize the likelihood of
interaction. This allows black hole events to be distinguished from
other standard model events. Detecting $TeV$ black hole formation
with UHECR detectors may be possible through the decay of
$\tau$-leptons generated by $\nu_\tau$'s that interact in the Earth
or in mountain ranges close to the detectors. A secondary $\tau$
generated through the decay of a BH has much less energy than the
standard model $\tau$ secondary. In addition, black holes may
produce multiple $\tau$-leptons in their evaporation, a unique
signature of $TeV$ gravity. Standard model processes that generate
multiple $\tau$-leptons are highly unlikely, the detection of
multiple $\tau$'s in earth-skimming and mountain crossing
neutrinos will be a smoking gun for black hole formation.\\

\section{Conclusion}
In this paper, using generalized uncertainty principle in a general
form as our primary input, we have calculated the temperature and
microcanonical entropy of a black hole in the framework of large
extra dimensional scenarios. Following results can be obtained from
our analysis:
\begin{itemize}
\item
In scenarios with extra dimensions, black hole temperature
increases(figure 1). This feature leads to faster decay and less
classical behaviors for black holes.
\item
It is evident that in extra dimensional scenarios final stage of
evaporation( black hole remnant) has mass more than its four
dimensional counterpart.
\item
In scenarios with extra dimensions, black hole entropy
decreases(figure 2). The classical picture breaks down since the
degrees of freedom of the black hole, i.e. its entropy, is small. In
this situation one can use the semiclassical entropy to measure the
validity of the semiclassical approximation.
\item
It is evident that in extra dimensional scenarios final stage of
evaporation( black hole remnant) has event horizon area more than
its four dimensional counterpart(figure 2).
\item
Black hole radiation is mainly on the brane. In other words, black
holes decay by emitting radiation mainly on the brane. This is in
accordance with the results of Emparan {\it et al}[21].
\item
Black hole production at the LHC and in cosmic rays may be one of
the early signatures of $TeV$-scale quantum gravity. Large samples
of black holes accessible by the LHC and the next generation of
colliders would allow for precision determination of the parameters
of the bulk space and may even result in the discovery of new
particles in the black hole evaporation. Limited samples of black
hole events may be observed in ultra-high-energy cosmic ray
experiments, even before the LHC era. If large extra dimensions are
realized in nature, the production and detailed studies of black
holes in the lab are just few years away. That would mark an
exciting transition for astroparticle physics: its true unification
with cosmology the Grand Unification to live for.

\end{itemize}
Therefore, in the framework of GUP, it seems that quantum black
holes are hotter, shorter-lived and tend to evaporate less than
classical black holes. Higher dimensional black hole remnants have
less classical features than four dimensional black holes. It is
evident from our calculations that black holes radiate mainly on the
four-dimensional brane-world.\\

\begin{figure}[ht]
\begin{center}
\includegraphics{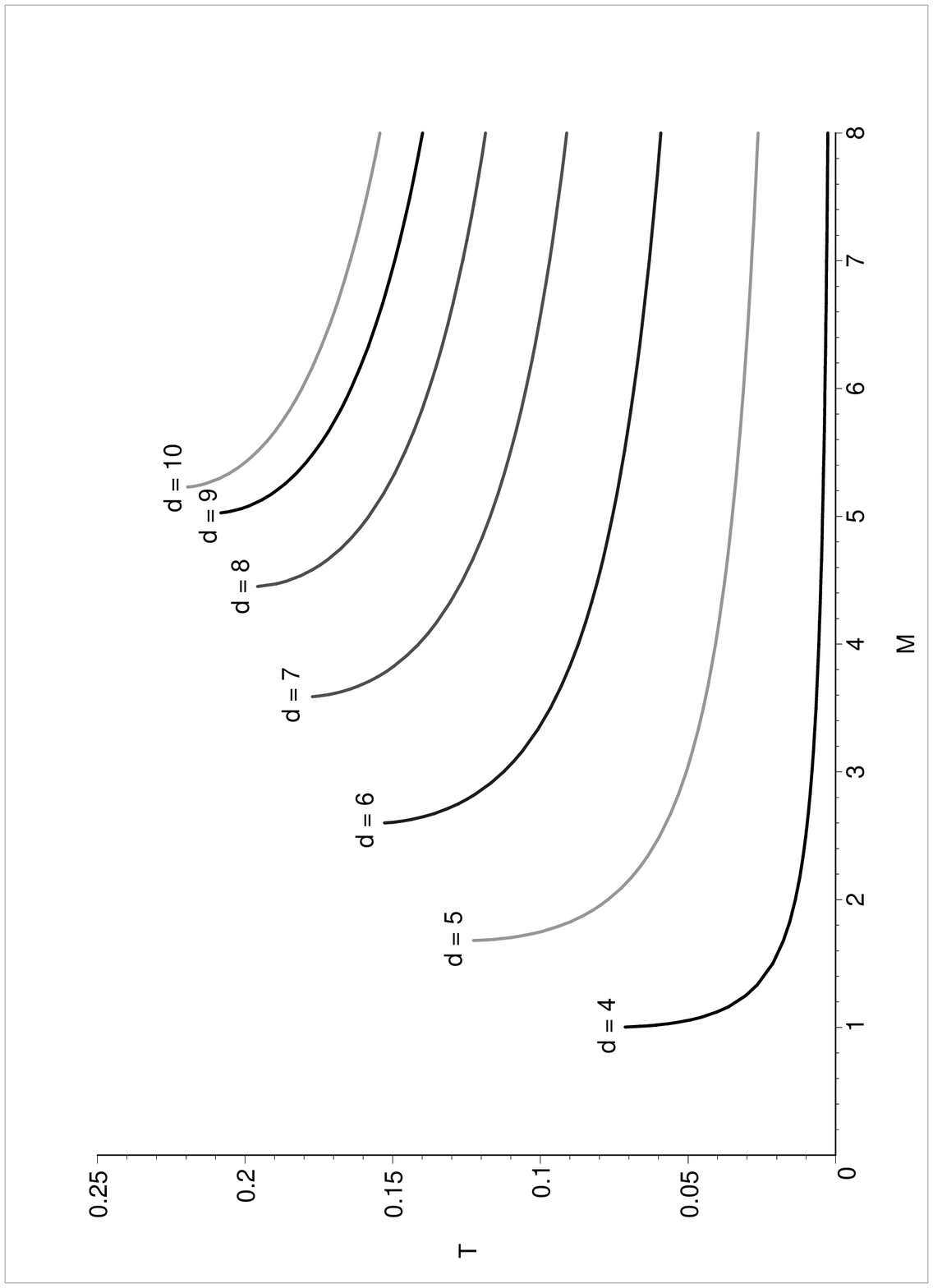}
\end{center}
\vspace{16 cm} \caption{\small {Temperature of black hole Versus its
mass in different spacetime dimensions. }} \label{fig:1}
\end{figure}
\begin{figure}[ht]
\begin{center}
\includegraphics{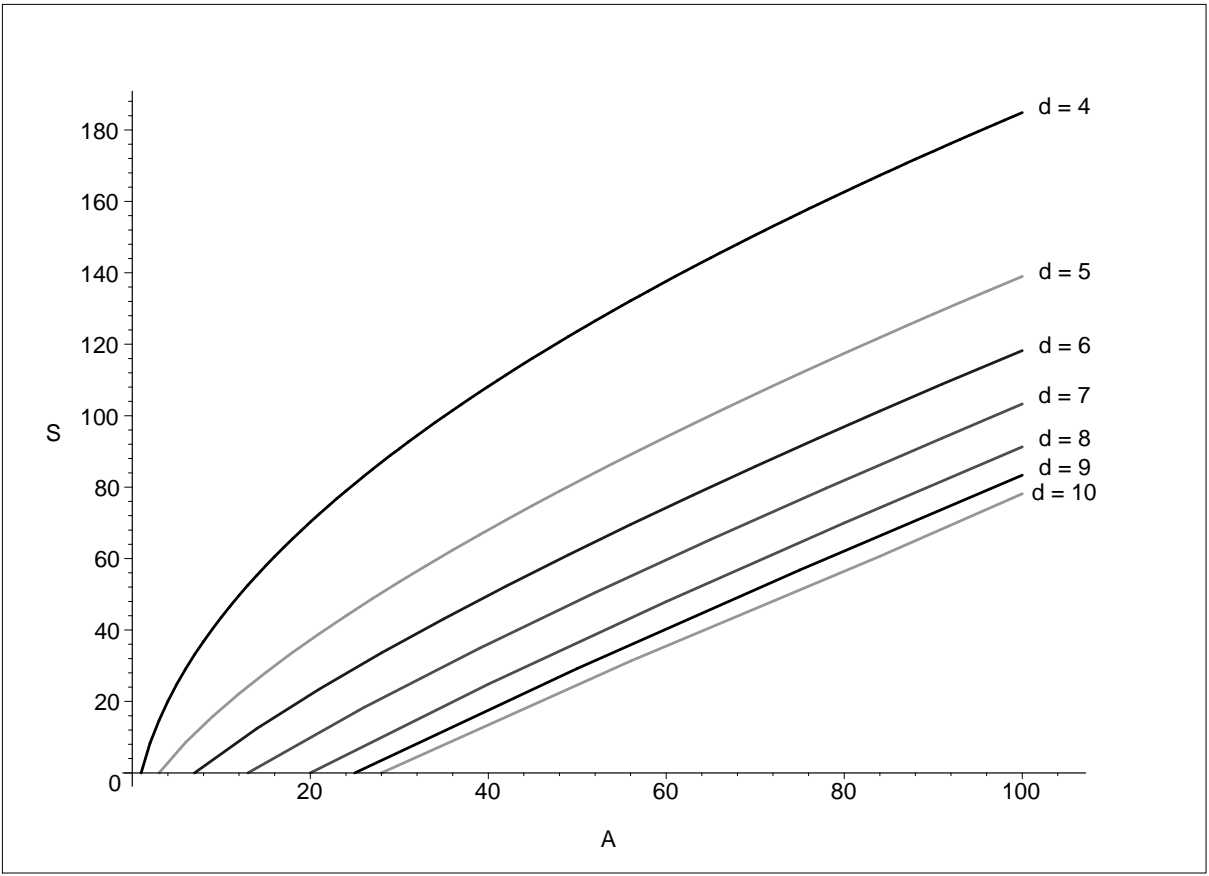}
\end{center}
\vspace{16 cm} \caption{\small {Entropy of black hole versus the
area of its event Horizon in different spacetime dimensions. }}
\label{fig:2}
\end{figure}

\end{document}